\documentclass{article}
\usepackage{amsfonts,amssymb,amsmath,euscript,xspace,cite}
\usepackage{amsthm,times}
\usepackage{a4}
\usepackage{graphicx}
\title{On models of a nondeterministic computation}
\author{M. N. Vyalyi\thanks{Dorodnitsyn Computing Center of RAS,
Vavilova, 40, Moscow, 119991, Russia. The work is supported by the
RFBR grants 08--01--00414, 05--01--02803--NTsNIL\_a and the grant NS
5294.2008.1.}}  
\date{\today}

\theoremstyle{plain}
\newtheorem{theorem}{Thorem}
\newtheorem{lemma}{Lemma}
\newtheorem{prop}{Proposition}
\newtheorem{cor}{Corollary}
\theoremstyle{definition}
\newtheorem{defi}{Definition} 
\newtheorem{rem}{Remark}

\let\al\alpha
\let\om\omega

\def\poly{\mathop{\mathrm{poly}}}
\def\lcm{\mathop{\mathrm{lcm}}}

\renewcommand*\P{\ensuremath{\mathrm {P}}}
\newcommand*\NP{\ensuremath{\mathrm {NP}}}
\newcommand*\PSPACE{\ensuremath{\mathrm {PSPACE}}}

\newcommand*\RE{\ensuremath{\mathrm {R.e}}}

\def\LogS{\ensuremath{\mathrm L}}
\def\NL{\ensuremath{\mathrm {NL}}}

\def\R#1{\ensuremath{#1\text{-}\mathrm {NFA}}}
\def\wR#1#2{\ensuremath{#2(#1)\text{-}\mathrm {NFA}}}
\def\dR#1{\ensuremath{#1\text{-}\mathrm {DFA}}}

\def\aDC{\ensuremath{\mathrm {Aux2DC}}}
\def\aNC{\ensuremath{\mathrm {Aux2NC}}}
\def\DR{\ensuremath{\mathrm {Aux2PDA}}}
\def\NR{\ensuremath{\mathrm {AuxN2PDA}}}

\def\ZZ{\mathbb Z}
\def\NN{\mathbb N}

\def\T{\ensuremath{\mathcal T}}

\def\wbeg{\triangleleft}
\def\wend{\triangleright}

\begin{document}
\maketitle

\begin{abstract} 
  In this paper we consider a nondeterministic computation by
  deterministic multi-head 2-way automata having a read-only access to an
  auxiliary memory. The memory contains additional data (a guess) and
  computation is successful iff it is successful for some memory
  content.

  Also we consider the case of restricted guesses in which a guess
  should satisfy some constraint.

  We show that the standard complexity classes such as \LogS, \NL, \P,
  \NP, \PSPACE{} can be characterized in terms of these models of
  nondeterministic computation. These characterizations differ from
  the well-known ones by absence of alternation.

\end{abstract}

The standard way to define a nondeterministic computation by an automaton
or a Turing machine is to change a transition function by a transition
relation. In a nondeterministic state of a computational device a
computation branches into several computation paths. 

There is another way to introduce a nondeterminism. Suppose that a
computational device has an additional data (a \emph{guess} or a
\emph{certificate} or a \emph{proof of correctness\/}) and performs a
deterministic computation operating with an input data and a guess
data.

Sometimes these variants of introducing nondeterminism lead to
equivalent computational models. The class \NP, for example, can be
defined in both ways using Turing machines. 

If we restrict computational power of a computational device these
variants may differ drastically. The aim of this paper is to
investigate models of nondeterminism based on the second variant for
multi-head 2-way automata. 

It is
well-known\footnote{O. H. Ibarra~\cite{Ibarra71} attributed this result to
A. Cobham and coauthors referring to an unpublished manuscript.} that
computation abilities of multi-head 2-way automata are equivalent to Turing
machines with a logarithmically bounded memory. So, they recognize
languages from the class $\LogS$. 

Nondeterministic (in the sense of transition relation) multi-head 2-way
automata recognize languages from the class $\NL$. One can rewrite a
definition of a nondeterministic automaton using the second way of
introducing nondeterminism. Let's imagine that a guess data are
written on an auxiliary  tape, which is 1-way read-only.
It is easy to see that using an 1-way guess tape leads
to an equivalent definition of a nondeterministic automaton. 

In this paper we consider a more general model of an auxiliary
read-only memory
(see definitions in Section~\ref{aam}).
Guess data are stored in cells of a  memory and at each moment
of time an automaton has an access to the exactly one memory
cell. Possible transitions between memory cells form a directed graph
(\emph{the memory graph\/}). An automaton can choose between finite
number of variants only. So, the natural condition on the memory graph
is a finite fan-out in each vertex (i.e. a memory cell). 

The most natural variant of the auxiliary memory is a 2-way tape. The
corresponding computational model appears to be very close to
nonerasing nondeterministic stack automata
(NENSA)~\cite{Ibarra71,HU79}. Similarly to multi-head NENSA, the
automata with 2-way read only guess tape recognize all languages from
the class \PSPACE.

It is possible to define in our settings a deterministic computation as
a specific case of a nondeterministic one. The deterministic automata
with 2-way guess tape are similar to nonerasing deterministic stack
automata (NEDSA) and also recognize the languages from the class
\PSPACE. 

We focus our attention on a more restricted memory model, so-called
1.5-way tape. It was used in research of quantum
automata~\cite{AmIw99}. For classic automata 1.5-way tape means an
1-way tape with an additional possibility to return into the first
cell.

The nondeterministic automata with 1.5-way tape also recognize the
class \PSPACE{} (Theorem~\ref{1.5=PSPACE} below). But deterministic
automata with this memory type recognize the class \P{} only
(Theorem~\ref{1.5det=P}). These results show that the 1.5-way guess tape is
potentially more suitable to characterize various complexity classes.

Also we introduce a nondeterministic computation
with a restricted guess. An example of restricted guess is a
\emph{sparse guess\/}. Sparseness of a guess means that a guess tape
contains the only one (or finitely many) non-empty symbol and the rest
symbols stored on the tape are empty. Using this model of a
nondeterministic computation gives the class \NP.

An interesting feature of all these results is a formal absence of
resource bounds in characterizations of resource-bounded classes such
as \P, \NP{} and so on. It should be noted that there is a primary
result of this sort: many heads are equivalent to logarithmic
space. The rest of results are based on this fact.

The main technical tool in study of the 1.5-way tape is calculations
modulo polynomially bounded integer. These calculations can be
performed on logarithmic space. To compute a length of a part of the
guess tape we use the simple algorithm: go along the part and increase
a counter modulo $p$. The latter operation can be done on logarithmic
space. The length can be restored from these data due to the Chinese
remainder theorem.

There are many results on characterizations of complexity classes in
terms of some sort of automata. The classes \LogS, \NL, \P, \PSPACE{}
have the well-known characterizations by deterministic,
nondeterministic, alternating and synchronized alternating 2-way
automata~\cite{alternation,sync-alt,Gef98}. There are also
characterizations of \NP{}, the polynomial hierarchy and some other
complexity classes in terms of alternating auxiliary stack
automata~\cite{HMcK03}. 

Our results differ from these characterization because the models
considered in this paper do not use alternation.

It is worth to mention a paper~\cite{Borchert03}, which contains
the characterizations of $\P$, $\NP$ and $\PSPACE$ in terms of
nondeterminism and so close to our results. The difference is in the
nature of nondeterminism introduced. In~\cite{Borchert03} 
nondeterministic colorings of $n$-dimensional words are
considered. Contrary, our main results concern the case of
1-dimensional guess memory.

The rest of paper is organized in the following way. In
Section~\ref{aam} we introduce our basic computational model:
multi-head 2-way automata with a nondeterministic auxiliary
memory. Section~\ref{nondetmemory} contains results about the 1-way,
the 1.5-way and the 2-way guess tapes.
In Section~\ref{restricted-guess} we introduce a model of a restricted
guess formally and give characterizations of \NP{} in terms of this model.
Section~\ref{final} contains some additional remarks on possible
variants of defining nondeterministic computation.

\section{Automata with an auxiliary memory}\label{aam}

In this section we provide definitions for a model of nondeterministic
computation by automata with an auxiliary read-only memory. The
definitions fix an informal idea explained in the
introductory section. They follow the  standard way of definition for
computational models.

\begin{defi}
A \emph{memory model} is a directed graph  $(M,E)$, the initial cell
$m_0\in M$ and a marking map $g\colon E\to G$ from the edges of the
graph to some finite set $G$. The marking map satisfies the following
conditions:
\begin{itemize}
\item $g(u,v)\ne g(u,w)$ for $v\ne w$ (different edges outgoing from
  the same vertex have different marks);
\item for each $u\in M$ and $a\in G$ there is an edge $(u,v)\in E$
  such that   $g(u,v)=a$. 
\end{itemize}
In other words, the map  $g$ restricted to the set of edges outgoing
from a vertex is a bijection.

For any finite alphabet $\Delta$ \emph{a memory content} $\mu$ is a map
$\mu\colon M\to \Delta$.
\end{defi}

\begin{defi}
An \emph{$h$-head automaton $A$ with an auxiliary memory of model $M$} is
characterized by
\begin{itemize}
\item a finite state set $Q$,
\item a finite input alphabet $\Sigma\cup\{\wbeg,\wend\}$, 
\item a finite memory alphabet $\Delta$, 
\item a transition function $\delta$, which maps a $(h+2)$-tuple 
  (the current state, symbols of the input word under the heads, the
  symbol in the current memory cell) to a $(h+2)$-tuple (a new state,
  a motion command for each head, a command of changing memory cell),
\item the initial state $q_0\in Q$,
\item the set of accepting states $Q_a\subset Q$.
\end{itemize}
Heads can move along the input words by one position per step.
So, a \emph{motion command} for a head is an element from the
set $\{-1,0,+1\}$. A \emph{command of changing  memory cell} is just an
element of the marking set  $G$ or an empty command (do not change the
cell). 
\end{defi}

An automaton $A$ operates on an input word $w\in\Sigma^*$ in natural
way. We assume that the input word is extended by markers
$\{\wbeg,\wend\}$ of the beginning and the end of the word. The
automaton starts from the initial state $q_0$, the initial position
of each head is the leftmost symbol of the input word, the initial
memory cell is $m_0$. The automaton applies the transition function on
each step of operation to modify its state, head positions and a
memory cell. For a fixed content of the auxiliary memory it generates
a sequence of configurations. The automaton stops iff it
reaches an accepting state.

\begin{defi}
The automaton $A$ \emph{accepts} the input word $w$ iff for some memory
content $\mu$ it stops an operation.

The automaton \emph{recognizes} the language $L$ iff for any $w\in L$ it
accepts $w$ and for any $w\notin L$ it do not accept $w$.
\end{defi}

We denote by $\R{M}$ the class of languages recognized by automata with
an auxiliary memory of model $M$. We denote by $\R{M}(h)$ the subclass of
languages recognized by automata with $h$ heads.

\subsection{Determinization}

As a specific case of a nondeterministic memory one can regard
deterministic automata equipped with a WORM-memory (write once, read
many). Such an automaton should fill a new memory cell by a symbol
when it enter the cell the first time. In further operation it can not
change the cell. Let's introduce a formal definition suitable for our purposes.

\begin{defi}
 A \emph{WORM-memory automaton on memory model $M$}
is characterized by
\begin{itemize}
\item a finite state set $Q$,
\item a finite input alphabet $\Sigma\cup\{\wbeg,\wend\}$, 
\item a finite memory alphabet $\Delta\cup\{\text{\sf void}\}$, 
\item a transition function $\delta$, which maps a $(h+2)$-tuple 
  (the current state, symbols of the input word under the heads, the
  symbol in the current memory cell) to a $(h+2)$-tuple (a new state,
  a motion command for each head, a command of changing memory cell),
\item the initial state $q_0\in Q$,
\item the set of accepting states $Q_a\subset Q$,
\item the set of writing states $Q_w\subset Q$.
\item a filling memory function $\varphi\colon Q_f\to \Delta$,
\end{itemize}
\end{defi}

At the start of operation all memory cells are void.  A WORM-memory
automaton operates in the same way as a nondeterministic $M$-automaton
except the moments of entering a writing state.
In that moment the filling function is applied to the
current state of the automaton. If the current memory cell is visited
at first time then the value of the filling function is assigned to
the cell and the automaton continues operation by application of the
transition function. An attempt to change the content of a cell visited before
causes the error as well as an attempt to apply the transition
function being at a void cell. In the case of an error the
automaton stops the operation and do not accept the input word. 

So, during a successful operation 
the automaton  enters a new memory cell in a writing state.
Also note
that if the automaton  writes the non-void symbol $d$ to the cell containing
the symbol $d$ then no error occurs. We call this property `a freedom of
writing the same'.

We denote by $\dR{M}$ the class of languages recognized by deterministic
automata with an auxiliary WORM-memory of model $M$.

\begin{lemma}\label{DFAinNFA}
  $\dR{M}\subseteq\R{M}$.
\end{lemma}

\begin{proof}
  Let $A$ be a WORM-$M$ automaton recognizing the language $L$ and $Q$
  is the state set of $A$. The
  state set of a nondeterministic
  $M$-automaton $A'$ recognizing the language $L$ is $Q\cup\{r\}$,
  where $r$ is an additional rejecting state.  The transition function
  of $A'$ coincide with the transition function of $A$ except writing
  states and the rejecting state. In a writing state $q\in Q_w$ the
  automaton compares the content $d$ of the current memory cell with
  $\varphi(q)$. If $d=\varphi(q)$ then the value of transition
  function is the same as for the automaton $A$. Otherwise, the
  transition leads to the rejecting state. In the rejecting state the
  automaton do nothing and the rejecting state is absorbing.

  An operation of the WORM-$M$ automaton $A$ on an input word $w$ gives
  a partial memory content $\eta\colon M\to\Delta$ 
  for memory cells visited during the
  operation. We denote by $\T(A,w)$ the set of memory contents
  extending $\eta$. In other words, each memory content
  $\mu\in \T(A,w)$ has in each cell visited by $A$ during the
  operation on the word $w$ the symbol written by $A$.

  Let $w\in L$. The automaton $A'$ accepts the word $w$ on any memory
  content from the set $\T(A,w)$. Indeed, it operates exactly in the
  same way as $A$ on this memory content.

  Let $w\notin L$. Let's consider the cases of memory content for the
  nondeterministic automaton $A'$.

  1. $A'$ is operating on $\mu\in \T(A,w)$. In this case its operation is also
     the same as for $A$. Here we use the property of freedom of
     writing the same. So, $A'$ do not accept.

  2. $A'$ is operating on $\mu'\notin \T(A,w)$. In this case
     $\eta(m)\ne \mu'(m)$ for some memory cell $m$ visited by the
     automaton $A$. Following the operation of the $A$ choose the
     first such memory cell $m_1$. Before entering  $m_1$ operation
     $A$ and $A'$ is the same. When entering $m_1$ the automaton $A'$
     reads a symbol $d\ne \phi(q)$, where $q$ is the current state due
     to the choice of $m_1$. It means that $A'$ goes to the rejecting
     state and do not accept the word $w$ on the memory content $\mu'$.
\end{proof}

\section{Complexity classes recognized by automata with auxiliary
     tape memory}
\label{nondetmemory}

\subsection{1-way tape}

\begin{figure}[!h]
\noindent
\begin{minipage}{0.43\textwidth}
  \centerline{\includegraphics{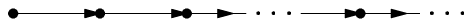}}
  \caption{1-way tape $W_1$}\label{pic-1way}
\end{minipage}
\hskip0.1\textwidth
\begin{minipage}{0.43\textwidth}
  \centerline{\includegraphics{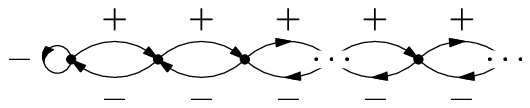}}
  \caption{2-way tape $W_2$}\label{pic-2way}
\end{minipage}
\end{figure}

Let $W_1$ be an infinite 1-way tape (Fig.~~\ref{pic-1way}).  The class
$\R{W_1}$ is just the class $\NL$. Indeed, a $W_1$-automaton can read
a symbol from the guess tape once. This symbol can be used to make a
nondeterministic choice in a transition relation for the case of the
standard definition of nondeterministic automaton.

Note also, that $\dR{W_1}=\LogS$ because we can simply ignore the
symbols written to the 1-way tape. 

\subsection{2-way tape}

Let  $W_2$ be an infinite 2-way tape
(Fig.~\ref{pic-2way}). For graphs of fan-out $>1$ we should also
indicate the marking of edges. In the case of $W_2$ the marking is
natural: mark `$+$' is placed on the edges going from a vertex $n$ to
the vertex $n+1$, mark `$-$' is placed on the edges going into the
opposite direction.

It was mentioned above that $\R{W_2}=\PSPACE$ because $W_2$-automata
is almost the same as nonerasing nondeterministic stack
automata and NENSA 
recognize the class \PSPACE~\cite{Ibarra71}.

The only difference between NENSA and $W_2$-automata is an ability of
NENSA to make arbitrary nondeterministic transitions while an
$W_2$-automaton should follow data read from the guess tape.  It means
that $W_2$-automata are weaker than NENSA, so
$\R{W_2}\subseteq\PSPACE$.  The reverse inclusion is valid even for
deterministic $W_2$-automata.  Indeed, a deterministic $W_2$-automaton
is able to write a computational history of a Turing machine
computation on a polynomially bounded space. For this purpose the
automaton should move on distances polynomially bounded by the input
size. But many heads are equivalent to logarithmic space and it is
easy to count polynomially many times using logarithmic memory.

Thus, $\R{W_2}\subseteq\PSPACE\subseteq\dR{W_2}\subseteq\R{W_2}$ (the
last inclusion is due to Lemma~\ref{DFAinNFA}).

\subsection{1.5-way tape}

\begin{figure}
  \centerline{\includegraphics{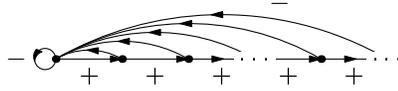}}
  \caption{1.5-way tape}\label{pic-1.5way}
\end{figure}

The memory model $W_{1.5}$ is pictured on the Fig.~\ref{pic-1.5way}.
Edges going to the right are marked by `$+$' and edges going to the
initial vertex are marked by `$-$'.

\begin{theorem}\label{1.5det=P}
  $\dR{W_{1.5}}=\P$.
\end{theorem}

We start from two simple observations.

\begin{lemma}\label{return-moves}
  Let $A$ be a $W_{1.5}$-automaton and $\#Q$ be the number of its
  states. Then any accepting computation of $A$ includes no more than
  $\#Q$ moves to the initial cell.
\end{lemma}
\begin{proof}
  After each return move the automaton $A$ scans the same tape
  content and its behavior is deterministic. So, if $A$ starts the
  scan process from the same state twice it loops and never reach an
  accepting state.

  Thus, the number of return moves is no more than the number of the states.
\end{proof}

\begin{prop}\label{fresh-cells}
   Let $A$ be a WORM-$W_{1.5}$ automaton, $k$ be the number of heads,
   $n$ be the length of the input word $w$ and $\#Q$ is the number of the
   states of $A$. If  $A$ accepts $w$ then between two subsequent
   return moves the automaton visits no more than $n^k\#Q$ new cells.
\end{prop}
\begin{proof}
  There are no more than $n^k\#Q$ surface configurations of
  $A$. Surface configurations are tuples (state, positions of heads).
  If the automaton pass through more than $n^k\#Q$ new cells, some surface
  configuration occurs twice. It means that the automaton loops and
  moves to the right infinitely.
\end{proof}

Now we are ready to prove Theorem~\ref{1.5det=P}.

\begin{proof}[Proof of Theorem~\ref{1.5det=P}]
  The inclusion $\P\subseteq \dR{W_{1.5}}$ follows from the fact that
  a WORM-$W_{1.5}$ automaton is able to simulate a WORM-$W_2$
  automaton on a polynomially bounded part of the memory tape. For
  this purpose one can use a polynomially bounded counter keeping the
  index of the current position on the guess tape.  When the
  $W_2$-automaton goes to the left, the simulating $W_{1.5}$-automaton
  returns to the initial position along the `$-$' marked edge and makes
  the required number of steps to the right according to the value of
  the position counter.

  In this way WORM-$W_{1.5}$ automaton can write down a computational
  history of a deterministic Turing machine computation polynomially
  bounded in time.

  Now we prove the reverse inclusion. Let $L\in \dR{W_{1.5}}$, $A$ be
  a WORM-$W_{1.5}$ automaton recognizing $L$, $Q$ be the state set of
  $A$, $k$ be the number of heads, $n$ be the length of the input word
  $w$.

  It follows from Lemma~\ref{return-moves} and
  Proposition~\ref{fresh-cells} that an accepting computation of $A$
  uses no more than $n^k(\#Q)^2$ cells. So, the automaton works on
  polynomially bounded auxiliary read only tape. It means that the
  total number of steps in an accepting computation is also
  polynomially bounded. It does not exceed $n^k\cdot n^k(\#Q)^2$. 

  Polynomially bounded in space and time computation of
  $W_{1.5}$-automaton can be simulated in polynomial time.
\end{proof}

Theorem~\ref{1.5det=P} shows that deterministic $W_{1.5}$-automata are
much weaker than deterministic $W_2$-automata. As for nondeterministic
automata, 1.5-way tape provides the same computational power as 2-way tape.

\begin{theorem}\label{1.5=PSPACE}
  $\R{W_{1.5}}=\PSPACE$.
\end{theorem}

\begin{proof}
  The statement is obvious in one direction: $\R{W_{1.5}}\subseteq
  \R{W_{2}}=\PSPACE$.

  To prove the reverse inclusion we show that a $W_{1.5}$-automaton is
  able to check correctness of a computational history for a Turing
  machine computation on a polynomially-bounded space.

  Without loss of generality we assume that the machine uses the
  binary alphabet $\{0,1\}$. Recall that a computational history is a
  sequence of a Turing machine configurations. A configuration is a
  word of form $\ell qa r$, where $\ell$ is the tape content  to
  the left of the head position, $q$ is the state of the machine, $a$
  is a currently read symbol, and $r$ is the tape content to the right
  of the head position.

  It is convenient to fix a length of a configuration. That is possible
  because we simulate a space bounded computation. For a computation
  on a space $s$ it is sufficient to deal with configurations of
  length $2s$.

  Each step of computation changes the configuration of  the
  machine. We will describe this change using arithmetic encoding of
  binary words~\cite{ShV-lang,Smallian}. Namely, a word
  $w\in\{0,1\}^*$ is encoded by a positive integer $c(w)$ written in binary as
  $1w$. 

  We will encode a configuration $\ell qa r$ by a 4-tuple
  $(c(\ell),q,a,c(r^R))$, where $r^R$ denote the word $r$ is the
  reversal of the word $r$.

  Changes of these data during a computation step are represented in
  the following table:
  \begin{center}\def\arraystretch{1.095}
    \begin{tabular}{cccc}
      \hline
      \multicolumn{4}{c}{Left move}\\
      \hline
      $c(\ell)$&$q$&$a$&$c(r^R)$\\
      \multicolumn{4}{c}{changes to}\\
      $\lfloor{}c(\ell)/2\rfloor$&$q'$&$c(\ell)\bmod2$&
      $a+2c(r^R)$\\[5pt]
      \hline
      \multicolumn{4}{c}{Right move}\\
      \hline
      $c(\ell)$&$q$&$a$&$c(r^R)$\\
      \multicolumn{4}{c}{changes to}\\
      $a+2c(\ell)$&$q'$&$c(r^R)\bmod2$&
      $\lfloor{}c(r^R)/2\rfloor$
    \end{tabular}
  \end{center}

  It is clear from the table that correctness of a computational
  history in the arithmetic encoding is equivalent to very simple
  arithmetic relations between neighbor pairs of configurations in the
  history. Depending on the pair $q,a$ and parities of $c(\ell)$,
  $c(r^R)$ each relation has a form 
  \begin{equation}\label{rel}
    y=2x,\ y=2x+1,\ x=2y+1,\ x=2y,
  \end{equation}
  where $x$ is the old value and $y$ is the new value of $c(\ell)$ or
  $c(r^R)$.
  
  Recall that we consider a computational  history of a computation on
  a polynomially bounded space. So, $c(\ell)=2^{\poly(n)}$,
  $c(r^R)=2^{\poly(n)}$, where $n$ is the input length. Thus, the
  relations~\eqref{rel} can be verified by calculations modulo
  $1,2,\dots,m=\poly(n)$. This fact follows from the Chinese remainder
  theorem and the prime number theorem~\cite{BS96}. 
  
  Now we are ready to describe a $W_{1.5}$-automaton verifying a
  computational history on the input word $u$ using a space $s$. The
  automaton expects a guess in form
  \begin{equation}\label{comphistory}
    u(\ell_0)q_0a_0u(r_0)\#
    u(\ell_1)q_1a_1u(r_1)\#\dots\#
    u(\ell_t)q_ta_tu(r_t)\#\#\:,
  \end{equation}
  where $\ell_0=0^s$, $a_0r_0=w0^{s-|w|}$, $\ell_i$, $q_i$, $a_i$,
  $r_i$ are components of the $i$th  configuration in the
  computational history, $q_t$ is a final state of the simulated
  Turing machine. The function
  $u(\ell)$ is the unary encoding of the number $c(\ell)$,
  i.e. $u(\ell)=*^{c(\ell)}$, where $*$ is the special symbol.

  The automaton makes $m=\poly(s)$ stages of computation. On the $p$th
  stage it verifies relations modulo $p$. It should verify the
  correctness of the the first block of the guess and the
  relations~\eqref{rel}. 

  The correctness of the first block on the
  input word $w=w_1w_2\dots w_n$ means that $c(\ell_0)=2^{s+1}$,
  $a_0=w_1$ and $c(r_0)=2^{s+1}+\overline{w_nw_{n-1}\dots w_2}$. Note
  that the right hand sides of these equalities can be computed modulo
  $p$ on a logarithmic memory without using the guess tape. After that
  the automaton computes residues modulo $p$ for the lengths of
  $u(\ell_0)$ and $u(r_0)$ in natural way: go along a word and count
  modulo $p$.

  The relations~\eqref{rel} are verified in the same manner: the
  automaton keeps in its logarithmic memory residues modulo $p$ of
  lengths $u(\ell_i)$, $u(r_i)$ as well as $q_i$, $a_i$ and compares
  them to the data of $(i+1)$th block computing residues modulo $p$ in
  natural way.

  If all checks are passed successfully for each residue and the state
  $q_t$ is a final state of the Turing machine then the automaton
  accepts the word $w$. Otherwise, it rejects (say, moves to the right
  infinitely).

  It is clear from the construction that if the simulated Turing
  machine accepts the word $w$ then the automaton also accepts it. Now
  suppose that the automaton accepts a word~$w$. It means that there
  is a guess in form~\eqref{comphistory} such that the automaton
  accepts $w$ on this guess. Because all relations~\eqref{rel} are
  true and the first block corresponds to the input we conclude that
  taking lengths $u(\ell_i)$, $u(r_i)$ modulo $\lcm(1,2,\dots, m)$
  give us the valid accepting computational history on the input $w$.
\end{proof}

\begin{rem}
  A simulation in Theorem~\ref{1.5=PSPACE} fails for a
  nondeterministic Turing machine computation. In the deterministic
  case there is the unique valid computational history for the
  computation on the input word~$w$. In the nondeterministic case
  there are several computation paths. So, correctness modulo small
  integers do no imply the total correctness. 
\end{rem}

\subsection{A perversed 1.5-way guess tape}

Let $\om$ be an infinite binary word. A modification $W_{1.5}^\om$ of
the 1.5-way guess tape differs from $W_{1.5}$ in the edge marking.
The mark `$+$' is assigned to the edges outgoing to the right from the
vertex $n$ iff  $\om_n=1$. The rest of edges are marked by `$-$'.

The marked graph $W_{1.5}^\om$ bears an information about the word
$\om$. There are continually many infinite binary words. So, it is
natural to expect non-decidable languages in some classes
$\R{W_{1.5}^\om}$. We present an example in the next theorem.

\begin{theorem}
  Let $L$ be a tally language (all its words are $1^n$). Denote by
  $\om_L$ the infinite word such that $\om_{2n-1}=1$ and $\om_{2n}=1$
  iff $1^n\in L$. Then $L\in \R{W_{1.5}^\om}$.
\end{theorem}
\begin{proof}
  A~$W_{1.5}^\om$-automaton recognizing $L$ expects a guess in a
  special form: each vertex contains an information about the
  direction of the edge marked by `$+$' and the initial vertex has a
  special root label. 

  The automaton should be able to verify the correctness of the guess.
  The algorithm of guess verification for the first $2n$ vertices
  checks the root label in the initial vertex and after that it
  makes $2n$ moves `to the right' according to the instructions of the
  guess, then it makes `the return move' also following the
  instructions of the guess. If the root label appears on the last step
  only then the automaton adopts the guess. Otherwise, it reject the
  guess as well as the input.

  It is easy to see that the algorithm indeed adopts the guesses of
  the form described above because any wrong instruction leads the
  automaton to the initial cell.

  After verification step the automaton can move along the guess tape
  following the instructions of the guess. It accepts the input word
  $1^n$ iff $2n$th instruction do not lead to the initial cell.
\end{proof}

\section{The restricted guess case}\label{restricted-guess}

One can put a restriction on the form of a guess. In the proofs
above we already use this technique. In this Section we consider the
notion of nondeterminism that arises in the restricted guess settings.

\begin{defi}
Let $\T\subseteq \Delta^M$ be a subset of possible memory contents. We
say that an automaton $A$ \emph{accepts a word $w$ with a
  $\T$-restricted guess} iff it accepts $w$
working on some memory content $\mu$ from the set $\T$.
\end{defi}

We denote by $\wR{\T}{M}$ the corresponding class of languages
recognizable by $M$-automata with a $\T$-restricted guess.

Of course, in general $\wR{\T}{M}\not\subseteq\R{M}$. For example, let
$\T$ is the set of all valid computational histories of a Turing
machines. Then  $\wR{\T}{W_{2}}$ 
contains all recursively enumerable languages. Indeed, a $W_2$-automaton
can verify the correctness of the first block of the history and the
correctness of all local changes of the machine state and symbols
around it. It is sufficient by the definition of the restriction.

To guarantee the inclusion $\wR{\T}{M}\subseteq\R{M}$ it is
sufficient to construct an automaton $V$ that checks compatibility of
memory content $\eta$ in visited cells with the set
$\T$. Compatibility means that $\eta$ can be extended to some
$\tau\in\T$. Below we apply this idea in specific cases.

We are interested in restrictions that describe subclasses of
$\R{M}$. As an example of this kind of restriction we introduce
\emph{sparse guesses\/}.

\paragraph*{Sparse guess.} Suppose that $\Delta=\{0\}\cup\Delta'$. 
A~$k$-sparse guess contains no more than $k$ symbols from the $\Delta'$.

We denote  by $U_k$ the set of $k$-sparse guesses. 

Below we consider sparse guesses for tape memories.

\subsection{Sparse guesses for 1.5-way tape}

An informal idea of guess verification described above gives us in the
case of the 1.5-way guess tape  the following lemma.

\begin{lemma}
  $\wR{U_k}{W_{1.5}}\subseteq\R{W_{1.5}}$ for any $k$. 
\end{lemma}
\begin{proof}
  Let $L$ be a language recognized by a $W_{1.5}$-automaton $A$ with a
  $U_k$-guess. An automaton $A'$  recognizing $L$ with the
  unrestricted guess runs in parallel the automaton $A$ and a special
  \emph{verifying automaton} $V$. The automaton $V$ has rejecting
  states which are absorbing. If $V$ is in a rejecting state then $A'$
  rejects. Otherwise, it accepts if  $A$ accepts.

  The automaton $V$ do not move itself. It looks at memory cells
  passed in motion of the automaton $A$ and change its
  state. Informally, it keeps an information about the number of
  non-zero symbols to the left of the current position. So, the states
  of the $V$ are the set $\{0,1,\dots, k+1\}$. The state $k+1$ is
  rejecting and thus is absorbing. 

  At the start and after each return move the state $V$ is set to $0$
  (except the case of state $k+1$).
  After reading a non-zero symbol and passing to the right $V$ changes
  the state $i$ by $i+1$ provided $i\leq k$. 
  
  If $w\in L$ then the automaton $A$ accepts it on a guess
  $\tau\in U_k$. The automaton $A'$ is also accepts $w$ on a guess
  $\tau$ because the state $k+1$ of the automaton $V$ can not be reached.

  If $w\notin L$ then no $U_k$-guess can enforce the automaton $A$ to
  accept $w$. The same holds for $A'$ and $U_k$-guesses. Suppose now
  that $A'$ accepts on a guess $\tau\notin U_k$. By construction $A'$
  do not visit more than $k$ different cells filled by non-zero
  symbols (otherwise, the automaton $V$ rejects). Let $\eta $ be the
  memory content of cells visited by $A'$ during the accepting
  computation. Then $\eta$ can be extended to some memory content
  $\tau'\in U_k$. The automaton $A$ works on the $\mu'$ in the same
  way as $A'$. In particular, it accepts on this guess. So, $w\in L$
  and we come to a contradiction. Thus, $A'$ rejects on any guess.
\end{proof}

The following inclusions are proved along the same lines.

\begin{lemma}\label{1<=k}
$\wR{U_1}{W_{1.5}}\subseteq\wR{U_k}{W_{1.5}}$. 
\end{lemma}
\begin{proof}
  Let $L$ be a language recognized by a $W_{1.5}$-automaton $A$ with a
  $U_1$-guess. Now we construct for $k\geq2$ an automaton $A'$ that
  recognizes $L$ with $U_k$-guess. The automaton $A'$ runs in parallel
  $A$ and a verifying automaton $V$ counting the number of non-zero
  symbols read. The construction of $V$ is the same as in the proof of
  the previous lemma. But now the state $2$ is rejecting for $V$.

  If $w\in L$ then the automaton $A$ accepts it on a guess $\tau$. The
  automaton $A'$ is also accepts $w$ on a guess $\tau'$ such that it
  coincides with $\tau$ in cells visited by $A$.

  If $w\notin L$ then no $U_k$-guess can enforce the automaton $A$ to
  accept $w$. The same holds for $A'$: $A'$ works in
  the same way as $A$ until reading the second non-zero symbol in which
  case the $A'$ rejects.

  Thus, $L\in \wR{U_k}{W_{1.5}} $.
\end{proof}

Now we give a characterization of the classes $\wR{U_k}{W_{1.5}}$.

\begin{theorem}\label{NPW1.5}
   $\wR{U_k}{W_{1.5}}=\NP$ for  $k\geq1$.
\end{theorem}

The proof of Theorem~\ref{NPW1.5} is splitted naturally into two parts.

\begin{lemma}\label{NPinU1W1.5}
   $\NP\subseteq\wR{U_1}{W_{1.5}}$.
\end{lemma}
\begin{proof}
  Let $L$ be an $\NP$-language. It means that there is a
  (deterministic) Turing
  machine $M$ and a polynomial $p$ such that for any $w\in L$ there
  is a certificate $y$ of polynomial size in the length of $w$
  ($|y|=p(|w|)$) such that $M$ accepts the input pair $w,y$ and for
  any $w\notin L$ there are no such certificate.

  A history of computation of $M$ on the input pair $(w,y)$ can be
  verified by a multi-head 2-way automaton $V$ with the indexed access
  to the history. It means that $V$ is equipped by a logarithmically
  small query tape which is read/write. The automaton $V$ has a
  special query state. Entering this state $V$ sends a query to the
  storage containing a string and receives in answer the value of the
  $i$th symbol of the string, where $i$ is written in binary on the
  query tape.

  It is easy to see that using polynomially small counters the
  automaton can verify a computational history of  of polynomial size.

  Now we are going to simulate the indexed access by a $U_1$-guess. In
  other words, we construct a $W_{1.5}$-automaton $I$ such that for any
  sequence $b_1,\dots, b_m$, where $m=\poly(n)$ and $0\leq b_i<
  b=O(1)$, there is a $U_1$-guess $\xi$
  such that the automaton $I$ can restore $b_i$ operating on the guess
  $\xi$. 

  At first we note that using a space $s$ one can compute the $k$th
  prime number $p_k$ for $1\leq k\leq   2^{s/C}$, where $C$ is the absolute
  constant. Indeed, the check of primality of an integer  $n$ written in
  binary on the space $\log n$ can be done by use of $O(\log n)$
  additional memory (containing auxiliary counters). Thus, using one
  more counter to keep the number of the last prime found one can
  compute $p_k$ on the space $s\leq C\log p_k$. From the prime number
  theorem~\cite{BS96} we conclude that  $p_k\sim k\ln k$, hence,
  $\log p_k\sim \log k+C_1\log\log k$ and for sufficiently large $k$
  the computation can be done on space $s\sim (C+1)\log k$.

  The automaton $I$ works in the following way. To compute a value of
  $b_i$ it computes  $p_i$ on its own logarithmic memory. Then it
  starts a motion along the guess tape and counts modulo $p_i$. When
  it reaches the non-zero symbol it returns the current residue modulo
  $p_i$ as the value of $b_i$ if $b_i<b$. Otherwise, it rejects.

  The Chinese remainder theorem implies that for any sequence  $b_i$
  there is an integer $N$ such that $N\equiv b_i\pmod{p_i}$ for all
  $0\leq i\leq m$. So, $I$ returns correct values of $b_i$ on the
  guess $0^{N-1}10\dots$.

  The  $W_{1.5}$-automaton $R$ with a $U_1$-guess recognizing the
  language $L$ is combined from the automata $V$ and $I$. It
  substitute calls of $I$ instead of queries of $V$.

  By construction, if $w\in L$ then $R$ accepts it. Let $w\notin
  L$. Consider an operation of $R$ on the input $w$. Possible results
  of operation $I$ form a sequence $(b_i')$ and the $V$ part of the
  automaton $R$ verifies it as a valid computational history.  Thus,
  the automaton $R$ rejects because there are no accepting
  computation.
\end{proof}

\begin{lemma}\label{UkinNP}
  $\wR{U_k}{W_{1.5}}\subseteq \NP$ for any $k$.
\end{lemma}
\begin{proof}
  We should construct a nondeterministic polynomial time algorithm to
  verify that a $W_{1.5}$-automaton $A$ accepts an input word $w$ on
  some guess $\tau\in U_k$.

  From $A$ and $w$ we construct in deterministic polynomial time an
  auxiliary automaton $B$. The states of $B$ are surface
  configurations of $A$, i.e. $(h+1)$-tuples (a state of $A$, head
  positions). So the number of states of $B$ is polynomially
  bounded. The automaton $B$ moves along the 1.5-way guess tape in the same way
  as the automaton $A$ do on the input $w$ except steps that do not
  change a memory cell. Following along the transitions of the
  automaton $A$ one can determine the next `moving' step in polynomial
  time. The automaton $B$ jumps to this step immediately.
  Accepting states of $B$
  are surface configurations such that $A$ is in an accepting state.

  Hence the problem is reduced to verification that there is a
  $U_k$-guess such that $B$ accepts on this
  guess. For this purpose we need the following claim.

\medskip

\textbf{Claim 1.} If $B$ accepts on some $U_k$-guess then it accepts on
a $U_k$-guess of exponential length.

\smallskip

  Consider an operation of $B$ on the guess
  $0^{x_0}s_00^{x_1}s_1\dots 0^{x_{k-1}}s_{k-1}0\dots$. Let $N$ be
  $\lcm(1,\dots,\#Q(B))$, where $\#Q(B)$ is the number of the states
  of $B$. Let's prove an intermediate claim.

\medskip

\textbf{Claim 2.} The operation of $B$ on the guess
$0^{y_0}s_00^{y_1}s_1\dots 0^{y_{k-1}}s_{k-1}0^{y_k}$ gives the same result
as the operation of $B$ on the guess $0^{x_0}s_00^{x_1}s_1\dots
0^{x_{\ell-1}}s_{k-1}0^{x_k}$ provided  $y_i\equiv x_i\pmod N$ for
$x_i>\#Q(B)$ and  $y_i=x_i$ for $x_i\leq \#Q(B)$.

\smallskip
  
  Indeed, a sequence of states of $B$ working on a part of the tape
  filled by zeroes is obtained by iterations of a map $\al_0\colon
  Q(B)\to Q(B)$. After $\leq \#Q(B)$ iterations the sequence
  $\al^n(q)$ became periodic.  The period depends on $q$ but in any
  case it is a divisor of $N$. Claim~2 is proved.

  Now the Claim~1 follows from the bound
  $N<2^{(\#Q(B))^2}$. (Actually, the bound is more more tight.)  

  Note that the parameters $x_i$ of an exponentially bounded guess can
  be written in binary nondeterministically in polynomial time.

  To complete a proof we construct a (deterministic) polynomial time
  algorithm verifying that $B$ accepts on the guess with parameters
  $x_i$. 

  By Lemma~\ref{return-moves} there are no more than  $\#Q(B)$ return
  moves during an accepting operation of $B$. So, the algorithm can call a
  procedure $F$ that by a state $q$ determine the behavior of $B$
  starting from the initial cell: either it reaches an accepting state
  or it makes the return move to the state $q'$.

  This procedure can be constructed easily using calls of the simpler
  procedure $F_0$ answering the same question concerning a behavior of
  the automaton on the part of tape filled by zeroes. More exactly, an
  input of the procedure is an integer $x$ written in binary and a
  state $q\in Q(B)$. The procedure $F_0$ should output the result of
  operation in one of three following forms:
  \begin{itemize}
  \item[(a)] $B$ reaches an accepting state working on the part $0^x$ of the
    tape without return moves;
  \item[(b)] $B$ reaches a return state and goes to the initial cell in the
    state $q'$;
  \item[(c)] $B$ passes the part $0^x$ and leaves it in the state $q'$.
  \end{itemize}

  To answer these questions the procedure $F_0$ represents the map
  $\al_0$  in a Boolean matrix form and applies fast algorithm
  of matrix exponentiation. 

  Let $B'$ be a modified automaton such that all accepting and return
  states of $B$ are changed by absorbing states. Let $\al'$ be a
  Boolean matrix of $\al_0$ for the automaton $B'$: $(\al')_{q'q''}=1$
  iff $\al_0(q')=q''$.

  The Boolean matrix multiplication is defined similarly to the usual
  matrix multiplication but addition and multiplication are changed by
  disjunction and conjunction respectively.

  The Boolean multiplication is associative due to distributive law
  for disjunction and conjunction. So, a Boolean power $(\al')^n$ can
  be computed in time $\poly(\log n)$ in usual way: by writing binary
  representation of $n$ and using subsequent squaring. Let $q'$ be an
  accepting or return state. Then it can be easily verified by a
  straightforward induction that
  \begin{itemize}
  \item $((\al')^n)_{qq'}=0$ if $q'$ is not reached during the
    operation of $B'$ on the string $0^n$,
  \item $((\al')^n)_{qq'}=1$ if $q'$ is  reached during the
    operation of $B'$ on the string $0^m$, where $m\leq n$. 
  \end{itemize}
  
  Computing Boolean powers of $\al'$ helps to choose between the
  above variants (a)--(c). 

  Indeed, if $((\al')^x)_{qq'}=0$ for each
  accepting or return state then we have the variant (c). The state
  $q'$ in question is in this case the only state such that
  $((\al')^x)_{qq'}=1$. 

  Otherwise, some accepting or return state is reached within the
  region $0^x$. To determine the state we apply a binary search to
  find out the smallest $n$ such that  $((\al')^n)_{qq'}=1$ for some
  accepting or return state $q'$. Looking at the state $q'$ we can easily
  distinguish the variants (a) and (b) and compute the  data required
  in each case. 
\end{proof}

\begin{proof}[Proof of Theorem~\ref{NPW1.5}]
  From Lemmata~\ref{1<=k},~\ref{NPinU1W1.5},~\ref{UkinNP} we conclude
  that 
$$
  \NP\subseteq\wR{U_1}{W_{1.5}}\subseteq
  \wR{U_k}{W_{1.5}}\subseteq\NP.
$$
\end{proof}

\begin{rem}
  In similar way it is possible to determine the result of operation
  of a  $W_{1.5}$-automaton on a guess containing polynomially many
  non-zero symbols. 
\end{rem}

\subsection{Sparse guesses for 2-way tape}

\begin{lemma}\label{UkW2}
    $\wR{U_k}{W_{2}}\subseteq\R{W_{2}}$ for any $k$. 
\end{lemma}

\begin{proof}[Sketch of proof]
  The idea is the same as for Lemma~\ref{1<=k}. We use a combined
  automaton that runs in parallel the recognizing and the verifying
  automata. The latter should be modified to include the moves to the
  left. The modification is straightforward.
\end{proof}

The class $\wR{U_1}{W_2}$ is rather weak. The reason is the absence of
the root label in the initial cell. Using a non-zero symbol as the root
label we obtain a subclass of $\wR{U_1}{W_2}$ that coincides with the
class $\aDC$ of languages recognized by deterministic 2-way counter
automata with a logarithmic auxiliary memory. The inclusion
$\aDC\subset\P$ follows from the Cook theorem~\cite{Cook71}. The Cook
theorem claims that $$\DR=\NR=\P,$$ where $\DR$ is the class of
languages recognized by deterministic 2-way pushdown automata with
a logarithmic auxiliary memory and $\NR$ is the class of languages
recognized by nondeterministic 2-way pushdown automata with a
logarithmic auxiliary memory.

To upperbound the class $\wR{U_1}{W_2}$ we state a rather obvious
proposition. 

\begin{prop}\label{trajectory}
  A trajectory of motion of a 2-way automaton $B$ along the tape filled by
  zeroes either became periodic with the period width bounded by 
  $\#Q(B)$, where $\#Q(B)$ is the number of the states of $B$, or 
  is an infinite repetition of right shifts by a distance $s$ along
  periodically repeated route. Here  $s\leq Q(B)$.
\end{prop}
\begin{proof}
  After $t\leq Q(B)$ steps a sequence of states became periodic. From
  this moment of time one of variants listed in the proposition became
  true. 
\end{proof}

\begin{theorem}
  $\wR{U_1}{W_2}\subseteq \aNC\subset\P$, where $\aNC$ is the class of
  languages recognized by nondeterministic 2-way counter automata
  with a logarithmic auxiliary memory.
\end{theorem}
\begin{proof}
  Let $L\in\wR{U_1}{W_2}$ is recognized by a $W_2$-automaton~$A$ with
  an $U_1$-guess. 

  Proposition~\ref{trajectory} implies that if  $A$ accepts the input
  word $w$ on some $U_1$-guess then it accepts the word $w$ on a guess
  such that a non-zero symbol is placed at polynomially bounded
  distance from the initial cell. (Look at the behavior of the
  automaton after visiting the non-zero symbol the first time.)

  The auxiliary counter automaton $B$ guesses nondeterministically
  the distance between the initial cell and the cell containing the
  non-zero symbol and keeps it in its logarithmic auxiliary
  memory. After that $B$ simulates an operation of $A$. The counter helps
  to simulate a behavior of $A$ when $A$ is to the right of the
  non-zero symbol. For the rest moments of time $B $ simulates the
  behavior of $A$ using the auxiliary memory. It keeps a polynomially
  bounded counter indicating the position of $A$ on the guess tape to
  the left of the non-zero symbol.

  The second inclusion in theorem follows from the Cook theorem
  mentioned above.
\end{proof}

Theorem~\ref{NPW1.5} implies that  $\wR{U_2}{W_2}\supseteq\NP$ because
two non-zero symbols can be used to mark the initial cell and provide
a $U_1$-guess for a $W_{1.5}$-automaton. The latter can be simulated
by a  $W_2$-automaton working on a guess of this kind.

Using Proposition~\ref{trajectory} one can prove the reverse
statement. The proof is similar to the proof of Lemma~\ref{UkinNP}. An
arbitrary guess is replaced by an exponentially bounded guess. After
that one can develop an algorithm computing the result of operation on
an exponentially bounded guess represented by parameters $x_i$ as in
the proof of Lemma~\ref{UkinNP}. So, we came to the theorem

\begin{theorem}
  $\wR{U_k}{W_2}=\NP$ for  $k\geq2$.
\end{theorem}

\section{Some other memory models and variants of nondeterminism}
\label{final}

In this final section we briefly outline several interesting
variants of memory models and possible extensions of definitions.

\subsection{Monoid memory}

Let $G$ be a monoid generated by a set $G'=\{g_1,\dots, g_n\}$.
Then the memory of type $(G,G')$ is defined by the Cayley
graph of the monoid $M$: the vertex set is $G$, an edge marked $g_k$
goes from a vertex $x$ to the vertex $xg_k$.

1-way and 2-way tapes are examples of monoid memory. It follows
immediately from definitions that $\R{W_1}=\R{(\NN,\{+1\})}$. Also it
is easy to see that $\R{W_2}=\R{(\ZZ,\{+1,-1)\})}$. For the inclusion
$\R{W_2}\supseteq\R{(\ZZ,\{+1,-1)\})}$ one should apply a useful trick
converting a tape infinite in both directions to a tape infinite to one
direction. For the reverse inclusion it is useful to use a root
labeling.  Walking around $\ZZ$, an automaton is able to check that
there is the only one vertex labeled as the root in the region
visited. 

There is a weak upper bound for an arbitrary monoid memory.

\begin{theorem}\label{recursive-upperbound}
  Let $M$ be a monoid. If the word problem for $M$ is decidable then
  $\R{M}\subseteq \RE$, where $\RE$ is the class of recursively
  enumerable languages.
\end{theorem}

Let make some general remarks before explaining the proof of the theorem.

From an $M$-automaton $A$ one can construct in polynomial time an
automaton $B$ with polynomially many states that walks on $M$ in the
same way as $A$ do. The
construction is in fact described in the proof of Lemma~\ref{UkinNP}.  

\begin{defi}
  An $M$-walking automaton $B$ is called \emph{halting} iff it reaches
  an accepting state on some memory content. 
\end{defi}

Let $S_M$ be a language consisting of descriptions of halting
$M$-automata. 

An upper bound on the class $\R{M}$ follows from the fact.

\begin{prop}\label{S_M}
  Any language $L\in\R{M}$ is polynomially reducible to $S_M$.
\end{prop}

The proof of this proposition repeats the argument from the proof of
Lemma~\ref{UkinNP}. 

The halting problem for an $M$-automata $B$ is in fact a problem of
conditional reachability in the state graph of $B$. Correctness
conditions stem from the fact that if the automaton comes to the same
cell of memory it should follow the same guess symbol stored in the
cell. In other words, a route $q_0,q_1,q_2,\dots $ along the state
graph of $B$ induces a route $m_0, m_1, \dots$ along the memory graph. The
route is correct if all transitions in moments corresponding to the
same cell $m$ go along the edges with the same mark $d\in\Delta$.

Let put this more formally. For any route $q_0,q_1,\dots$ along the
state graph the corresponding route $m_0, m_1, \dots$ along the memory
graph introduces an equivalence relation between positions in the
route: $i\sim_M j$ iff $m_i=m_j$. On the other hand, the route
$q_0,q_1,\dots$  determines the word $\xi$ in the alphabet $G\times \Delta$
of form $(g_1,d_1),(g_2,d_2),\dots$, where $g_i$, $d_i$ are memory
edge mark and memory symbol corresponding to the step $i$.

Define a language $L(B)$ as the language of words $\tau\in
(G\times\Delta)^*$ generated by routes from the start state to some
accepting state. By definition the language $L(B)$ is regular.
The halting words are in the language $L(B)$. A~word is  halting iff
$d_i=d_j$ for all $i,j$ such that $i\sim_Mj$.

\begin{proof}[Sketch of proof of Theorem~\ref{recursive-upperbound}]
  Let prove that $S_M$ is recursively enumerable.

  Since the word problem for $M$ is decidable there is an algorithm
  computing for each route from $(G\times \Delta)^*$ the equivalence
  relation $\sim_M$. 

  To enumerate halting automata the enumeration algorithm starts an
  enumeration of all pairs $(B,\xi) $, $\xi\in L(B)$. For each pair
  the algorithm computes the relation $\sim_M$ and checks the
  correctness conditions. If the conditions hold then the algorithm
  outputs~$B$.

  Application of Proposition~\ref{S_M} completes the proof.
\end{proof}

For many monoids and groups the bound of
Theorem~\ref{recursive-upperbound} is exact.

\subsection{$\ZZ^2$ memory}

The generators of $\ZZ^2$ are chosen naturally: $(\pm1,0)$ and
$(0,\pm1)$.

\begin{theorem}
  $\R{ZZ^2}=\RE$.
\end{theorem}
\begin{proof}[Sketch of proof]
  The word problem for $\ZZ^2$ is decidable. So, by
  Theorem~\ref{recursive-upperbound} $\R{ZZ^2}\subseteq\RE$. 

  On the other hand, a $\ZZ^2$-automaton is able to verify the
  correctness of computational history of an arbitrary Turing machine
  computation.  The automaton expects a guess containing subsequent
  Turing machine configurations in subsequent rows of
  $\ZZ^2$. Correctness of computational history in this form is a
  conjunction of local conditions that can be verified by the
  automaton walking on $\ZZ^2$.
\end{proof}

\begin{cor}
  Let $G$ be a group with decidable word problem and $\ZZ^2<G$.
  Then $\R{G}=\RE$.
\end{cor}

\subsection{Multi-head access to the guess data}

Our definitions permit a local access to the guess data.  Typically, a
ra;axation of this property leads to the class \RE{} of recursively
enumerable languages.

For example, if we allow two heads on 1-way tape we already get the
class $\RE$. Indeed, one can verify an arbitrary computational history
using two 1-way heads. 

Note that even for a sparse encoding two heads on the 2-way guess tape are too
much and we get \RE. Indeed, two parts of an arbitrary length can be
used to simulate an automaton with two counters. But such an automaton
is able to make an universal computation.

\subsection*{Acknowledgments}

The author is deeply indebted to S. Tarasov for numerous valuable
discussions and helpful references.

\end{document}